\documentclass[prl, notitlepage, showpacs, floatfix, reprint, superscriptaddress, reprint, nobalancelastpage]{revtex4-1}

\usepackage{amsmath}
\usepackage{amssymb}
\usepackage{amsthm}
\usepackage[export]{adjustbox}
\usepackage{amsfonts}
\usepackage{verbatim}
\usepackage{listings}
\usepackage{enumerate}
\usepackage{latexsym}
\usepackage{psfrag}
\usepackage{bm}
\usepackage{dsfont}
\usepackage[all]{xy}
\usepackage{graphicx}
\usepackage{color}
\usepackage{mathtools}
\usepackage{eufrak}
\usepackage[percent]{overpic}
\usepackage{epsfig,slashed}
\usepackage{epstopdf}
\usepackage{lipsum}
\usepackage{float}
\usepackage{mathtools}
\allowdisplaybreaks
\usepackage[colorlinks=true]{hyperref}
\hypersetup{
    bookmarks=true,         
    unicode=false,          
    pdftoolbar=true,        
    pdfmenubar=true,        
    pdffitwindow=false,     
    pdfstartview={FitH},    
    pdftitle={Enhanced thermal Hall effect in the square-lattice Neel state},    
    pdfauthor={Rhine Samajdar et al.},     
    pdfsubject={},   
    pdfcreator={},   
    pdfproducer={}, 
    pdfkeywords={} {} {}, 
    pdfnewwindow=true,      
    colorlinks=true,       
    linkcolor=blue, 
    citecolor=blue,        
    filecolor=blue,      
    urlcolor=blue           
}

\usepackage{setspace}
\usepackage{natbib}
\usepackage{subfigure}
\usepackage[mathscr]{euscript}
\usepackage[export]{adjustbox}
\usepackage[bottom]{footmisc}
\usepackage{textcomp}
\usepackage{braket}
\makeatletter
\newsavebox\myboxA
\newsavebox\myboxB
\newlength\mylenA

\newcommand*\xoverline[2][0.75]{%
    \sbox{\myboxA}{$\m@th#2$}%
    \setbox\myboxB\null
    \ht\myboxB=\ht\myboxA%
    \dp\myboxB=\dp\myboxA%
    \wd\myboxB=#1\wd\myboxA
    \sbox\myboxB{$\m@th\overline{\copy\myboxB}$}
    \setlength\mylenA{\the\wd\myboxA}
    \addtolength\mylenA{-\the\wd\myboxB}%
    \ifdim\wd\myboxB<\wd\myboxA%
       \rlap{\hskip 0.5\mylenA\usebox\myboxB}{\usebox\myboxA}%
    \else
        \hskip -0.5\mylenA\rlap{\usebox\myboxA}{\hskip 0.5\mylenA\usebox\myboxB}%
    \fi}
\makeatother

\usepackage{latexsym}
\DeclareMathOperator{\Tr}{Tr}
\def \beq{\begin{eqnarray}}
\def \eeq{\end{eqnarray}}
\def \ua{\uparrow}

\def \da{\downarrow}

\def \Q{{\bm Q}}
\def \S{{\bm S}}
\def \r{{\bm r}}

\def \k{{\mathbf{k}}}
\def \q{{\mathbf{q}}}

\newcommand{\nn}{\nonumber \\}

\newcommand{\pdagger}{{\phantom{\dagger}}}
\renewcommand{\vec}[1]{\boldsymbol{#1}}
\newcommand{\figref}[1]{Fig.~\ref{#1}}
\newcommand{\equref}[1]{Eq.~(\ref{#1})}

\DeclareGraphicsExtensions{.png}
\begin{document}
\title{Enhanced thermal Hall effect in the square-lattice N\'eel state}
\author{Rhine Samajdar}
\affiliation{Department of Physics, Harvard University, Cambridge, MA
02138, USA}
\author{Mathias S. Scheurer}
\affiliation{Department of Physics, Harvard University, Cambridge, MA
02138, USA}
\author{Shubhayu Chatterjee}
\affiliation{Department of Physics, University of California, Berkeley, CA 94720, USA}
\author{Haoyu Guo}
\affiliation{Department of Physics, Harvard University, Cambridge, MA 02138, USA}
\author{Cenke Xu}
\affiliation{Department of Physics, University of California, Santa Barbara, CA 93106, USA}
\author{Subir Sachdev}
\affiliation{Department of Physics, Harvard University, Cambridge, MA
02138, USA}
\date{\today \\
}

\begin{abstract}
Recent experiments on several cuprate compounds have identified an enhanced thermal Hall response in the pseudogap phase. Most strikingly, this enhancement persists even in the undoped system, which challenges our understanding of the insulating parent compounds. To explain these surprising observations, we study the quantum phase transition of a square-lattice antiferromagnet from a confining N\'eel state to a state with coexisting N\'eel and semion topological order. The transition is driven by an applied magnetic field and involves no change in the symmetry of the state. The critical point is described by a strongly-coupled conformal field theory with an emergent global $SO(3)$ symmetry. The field theory has four different formulations in terms of $SU(2)$ or $U(1)$ gauge theories, which are all related by dualities; we relate all four theories to the lattice degrees of freedom. We show how  proximity of the confining N\'eel state to the critical point can explain the enhanced thermal Hall effect seen in experiment.

\end{abstract}

\maketitle

The thermal Hall effect has attracted much attention in recent years as a powerful tool to gain information about the nature of excitations in exotic materials as, for instance, in the spin-liquid candidate system $\alpha$-RuCl$_3$ \cite{2018PhRvL.120u7205K}. Grissonnanche {\it et al.} \cite{Taillefer19} measured the thermal Hall effect in the normal state of four different copper-based superconductors. A strong signal is found starting from optimal doping, where the pseudogap phase ends, all the way to the insulating parent compounds. These observations are quite surprising, as the insulator is expected to be a conventional N\'eel state, and spin-wave theory shows that this state has a much smaller thermal Hall response in an applied magnetic field than that observed \cite{SCSS19}. There is no sign of a quantized thermal Hall response though, so the insulator is not in a state with topological order and protected edge excitations.

\begin{figure}[bt]
\begin{center}
\includegraphics[width=\linewidth]{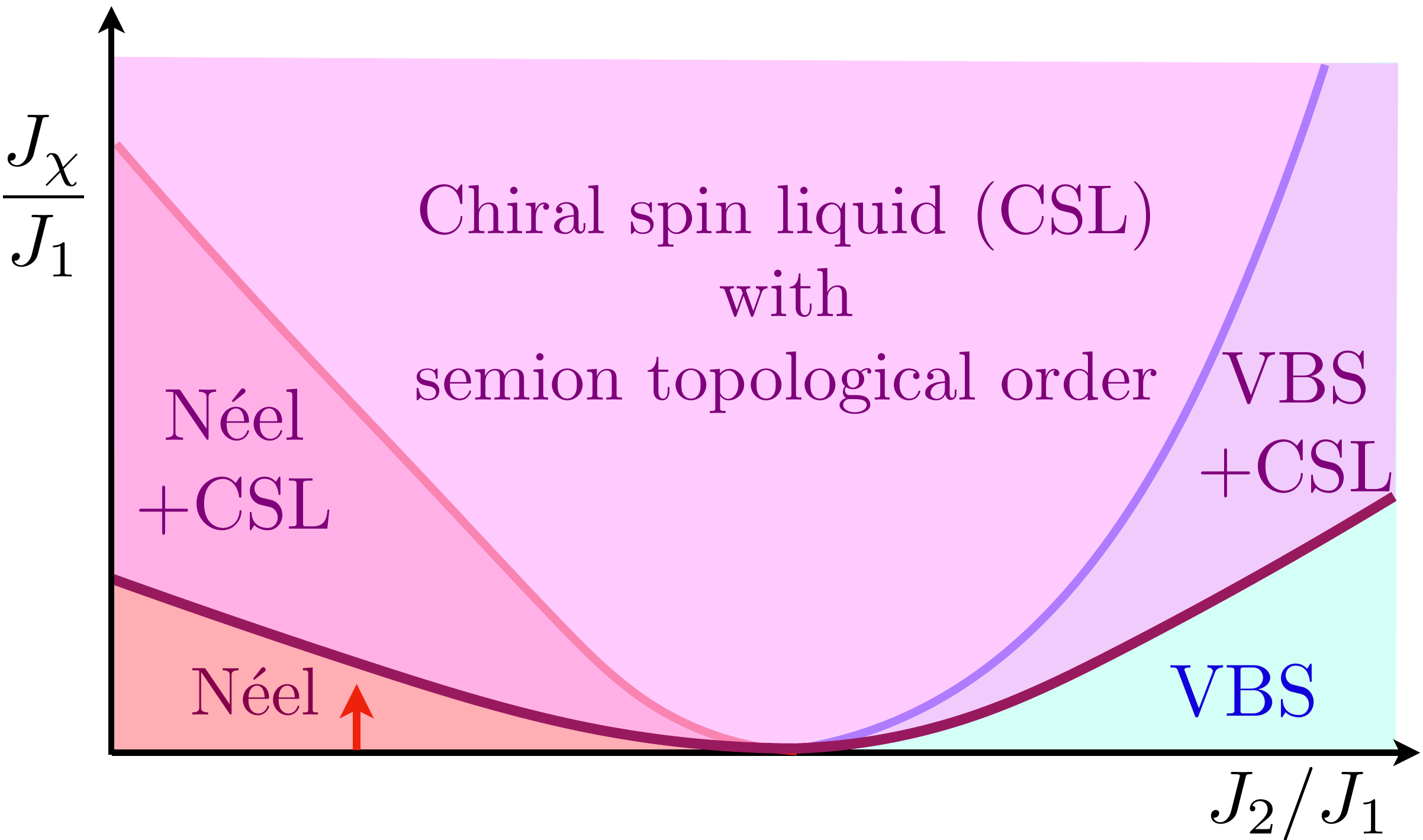}
\caption{\textbf{Proposed schematic phase diagram of $H_1 + H_B$ at ${\bm B}_Z=0$} (see Fig.~\ref{fig:MFPD}a for a phase diagram with nonzero ${\bm B}_Z$). By varying the first, $J_1$, and second, $J_2$, nearest neighbor exchange interactions and the orbital coupling $J_\chi$ in \equref{CFTco}, the antiferromagnet on the square lattice shows phases with combinations of N\'eel, valence bond solid (VBS), and chiral spin liquid topological order (CSL). The phase boundaries are presumed to meet at a $SO(5)$-symmetric (near) critical point at which $J_\chi$ is a relevant perturbation, and the phase boundaries all scale as $J_\chi \sim |J_2 - J_{2c}|^{\lambda_\chi/\lambda_2}$; we expect $\lambda_\chi/\lambda_2 > 1$. In this work, we imagine starting from the N\'eel state at zero magnetic field, $J_\chi=0$, close to the boundary of VBS order such that a small value of field-induced $J_\chi$ can already drive the system close to the phase boundary with N\'eel $+$ CSL (indicated by the red arrow). We note that the existence of a $SO(5)$ critical point is not a precondition for a continuous N\'eel to N\'eel $+$ CSL transition.
}
\label{fig:neelvbscsl}
\end{center}
\end{figure}

In this paper, we shall study the possibility that the orbital coupling of the applied magnetic field can drive the conventional, confining, N\'eel insulator to a state which has semion topological order \cite{Kalmeyer} coexisting with N\'eel order (see Fig.~\ref{fig:neelvbscsl}).
We assume that the current experiments are at a field where the ground state is a conventional N\'eel state whose only low energy excitations are spin waves, and we shall describe how the proximity to the lower quantum phase boundary in Fig.~\ref{fig:neelvbscsl} can enhance the thermal Hall response of such a conventional state.
The applied field and the N\'eel order break spin-rotation, time-reversal, and mirror-plane symmetries, and the states on both sides of the transition have an identical pattern of symmetries.
So the quantum phase transition only involves the onset of topological order.
We shall obtain the universal critical field theory describing the vicinity of the lower phase boundary in Fig.~\ref{fig:neelvbscsl} at low temperatures ($T$).

Remarkably, we find that the critical theory is one that has been carefully studied \cite{Benini17} in the context of the recent advances in dualities of non-Abelian conformal gauge theories in 2+1 spacetime dimensions \cite{Aharony15,Hsin16,Aharony16}.
The theory of interest has {\it four\/} different dual formulations in terms of relativistic field theories, and we will relate all of them to theories of the lattice antiferromagnet: the assumption of universality at the quantum phase transition then provides a new route to obtaining the dualities.

We are interested in spin $S=1/2$ antiferromagnets with spin operators ${\bm S}_i$ on the sites, $i$, of the square lattice, and Hamiltonian $H=H_1 + H_B$. The first term has the form
\beq
H_1 = \sum_{i<j} J_{ij} {\bm S}_i \cdot {\bm S}_j + \ldots,
\eeq
which describes near-neighbor exchange interactions and possible ring-exchange terms all of which preserve the global $SU(2)$ spin-rotation, time-reversal, and all square-lattice symmetries.
The second term, induced by the applied magnetic field, is
\beq\label{CFTco}
H_B = J_\chi \sum_{\triangle}  \S_i \cdot (\S_j \times \S_k) - \sum_i {\bm B}_Z \cdot {\bm S}_i\,.
\eeq
The $J_\chi$ term couples to the scalar spin chirality, and is induced by the orbital coupling of the applied magnetic field to the underlying electrons \cite{SenChitra}. It preserves lattice translations and rotations, but explicitly breaks time-reversal and mirror-plane symmetries while preserving their product. The value of $J_\chi$ itself is proportional to the small magnetic flux penetrating the square lattice. The second term in $H_B$ is the Zeeman term, and the electron magnetic moment has been absorbed in the definition of ${\bm B}_Z$. We do not include spin-orbit interactions; we note that with ${\bm B}_Z \neq 0$, spin-orbit interactions can enhance the stability of chiral topological phases similar to those discussed here \cite{Kitaev06}, and we do not expect such interactions to modify the universal critical theories presented below.

Numerical studies of $H$ at ${\bm B}_Z=0$ and $J_\chi \neq 0$ on the kagome \cite{Bauer14,YinChen15a,Sheng18} and triangular \cite{Sheng16,Lauchli17,McCulloch17,Sheng17} lattices have found convincing evidence, above very small values of $J_\chi$ (values as small as $J_\chi/J_1 = 0.0014$ in Fig.~19 of Ref.~\cite{McCulloch17}) for a `chiral spin liquid': a gapped state with semion topological order, but no antiferromagnetic order. More recently, a study \cite{Szasz18} of the Hubbard model on the triangular lattice found evidence for the same chiral spin liquid even at $J_\chi = 0$. On the square lattice, Nielsen {\it et al.\/} \cite{Cirac13} studied the antiferromagnet with first ($J_1$) and second ($J_2$) neighbor exchange and a nonzero $J_\chi$, and found evidence for the chiral spin liquid at quite small values of $J_\chi$, but in relatively small system sizes. These strong effects of a small $J_\chi$ can be understood by the proximity to a critical spin liquid at which an infinitesimal $J_\chi$ is a relevant perturbation.
The phase diagram we propose for the square lattice $J_1$-$J_2$-$J_\chi$ antiferromagnet is summarized in \figref{fig:neelvbscsl}, and the critical spin liquid is realized by the deconfined critical point at $J_\chi=0$ between the N\'eel and valence bond solid (VBS) states. Recent analyses \cite{Wang17} have shown that a relevant $J_\chi$ at this critical point does indeed lead to semion topological order. 
At such a critical point there is a discontinuous jump in the thermal Hall conductivity at low $T$ from $\kappa_{xy}/T = 0$ at $J_\chi = 0$ to $|\kappa_{xy}/T| =
(\pi/6)(k_B^2 /\hbar)$ \cite{Cappelli:2001mp} at infinitesimal $J_\chi$, and we will use proximity to this discontinuity to obtain the enhanced thermal Hall response in the N\'eel state.
We will show that turning on $J_\chi$ at values of $J_2/J_1$ smaller than at the deconfined critical point
({\it e.g.\/} along the red arrow in Fig.~\ref{fig:neelvbscsl}) leads to a state with coexisting N\'eel and semion topological order across a novel quantum critical phase boundary whose universal theory is obtained below. 
We refer to the Methods for further discussion of Fig.~\ref{fig:neelvbscsl}.

Our analysis starts from a model \cite{Wang17} of the square-lattice N\'eel state as the confining phase of a $SU(2)$ gauge theory of fluctuations about a `$\pi$-flux' mean-field state \cite{Affleck88}. In this formulation, the spins are represented by fermionic spinons $f_{i \alpha}$, ($\alpha = \uparrow, \downarrow$) via ${\bm S}_i = (1/2) f_{i \alpha}^\dagger {\bm \sigma}_{\alpha\beta} f_{i \beta}$, where ${\bm \sigma}$ are the Pauli matrices. This spinon representation induces a $SU(2)$ gauge symmetry \cite{AffleckAnderson}, and a full treatment requires careful consideration of the associated $SU(2)$ gauge field. However, much can be learnt from a mean-field theory in which we ignore the $SU(2)$ gauge fluctuations: we will analyze such a mean-field theory now, and turn to the gauge fluctuations later.

\begin{figure*}[tb]
\begin{center}
\includegraphics[width=0.90\textwidth]{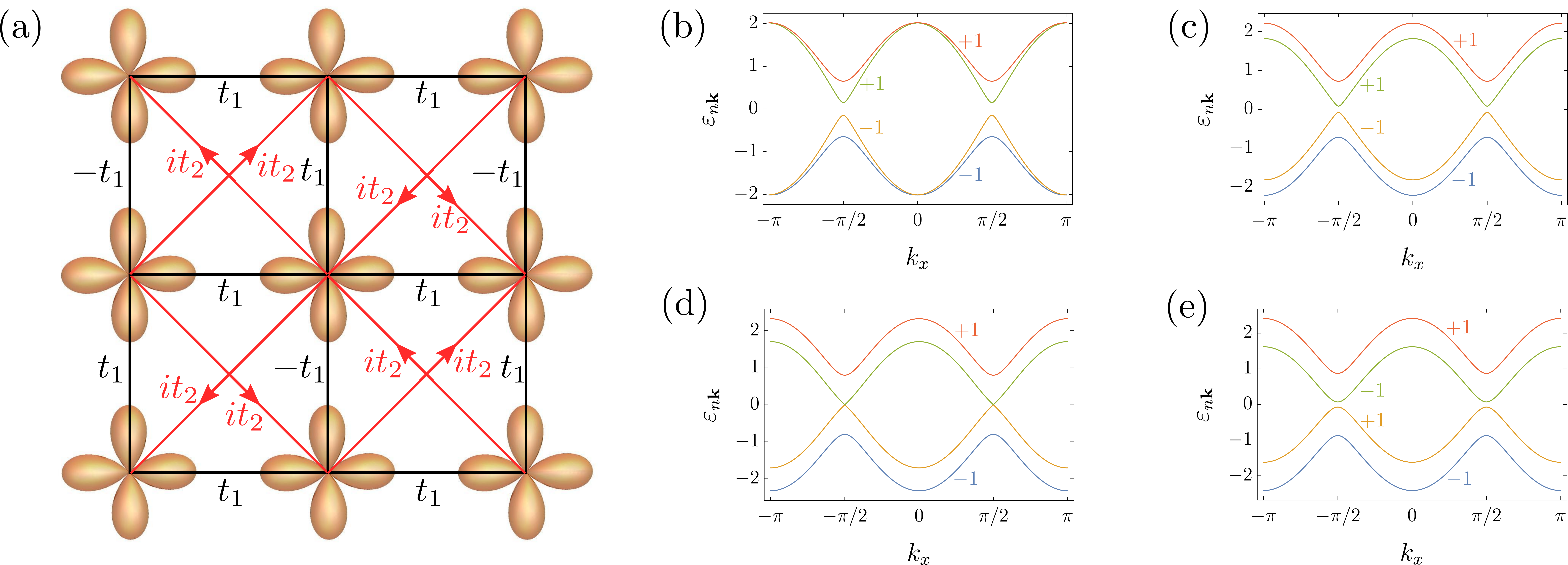}
\caption{\textbf{Ansatz and spectrum of spinon Hamiltonian.} \textbf{a}, The nearest ($t_1$, black) and second-nearest neighbor ($t_2$, red) hopping matrix elements for the spinon mean-field Hamiltonian in \equref{eq:fullH} on the square lattice formed by the Cu atoms (indicated in brown). The applied magnetic field induces a nonzero hopping $it_2$ and there is a uniform $\pi/2$ flux through each elementary triangle. In (\textbf{b}-\textbf{e}), we plot the evolution of the band structure of the Hamiltonian in Eq.~\eqref{eq:fullH} along the line $k_y = 0$, with ${\bm N} = 0.5 \hat{z}$, $t_1 = 1$, $t_2 = 0.10$, upon changing the Zeeman field, which is taken to be $|{\bm B}_Z| =$ (\textbf{b}) $0$, (\textbf{c}) $0.4 $, (\textbf{d}) $ B^{(c)}_Z \approx 0.6245$, and (\textbf{e}) $0.8$. The Chern numbers (indicated in the respective color) of the two lowest bands switch from $\{-1,-1\}$ to $\{-1,1\}$ as $|{\bm B}_Z|$ is increased across the phase boundary.}
\label{fig:fermionpi}
\end{center}
\end{figure*}

\section*{Mean-field theory}
After inserting the spinon representation of ${\bm S}_i$ in $H$, and a mean-field factorization respecting lattice and gauge symmetries, we obtain the quadratic spinon Hamiltonian \cite{WWZ89,WenPSG,Wang17,scheurer2018orbital}
\begin{eqnarray}
\label{eq:fullH}
H_f = &-& \sum_{i<j} \left( {t}^{}_{ij}  f_{i\alpha}^\dagger f^{\phantom{\dagger}}_{j\alpha} + {t}^\ast_{ij} f_{j\alpha}^\dagger f^{\phantom{\dagger}}_{i\alpha} \right) \label{eq:Hf}
\\ &-& \frac{1}{2} \sum_i \left( {\bm B}_Z + \eta^{}_i \, {\bm N} \right) \cdot f_{i \alpha}^\dagger {\bm \sigma}^{\pdagger}_{\alpha\beta} f^{\phantom{\dagger}}_{i \beta} \,. \nonumber
\end{eqnarray}
The pattern of the $t^{}_{ij}$ is shown in Fig.~\ref{fig:fermionpi}a.

The first-neighbor hopping, $t_1$,  arises from the factorization of the exchange couplings in $H_1$. The second-neighbor hopping, $\pm i t_2$, arises from the scalar spin chirality term $J_\chi$, and has the same symmetry as the orbital coupling of the underlying electrons to the magnetic field orthogonal to the plane of the square lattice. We have assumed a nonzero N\'eel order, and this leads to the ${\bm N}$ term after factorization of $H_1$; $\eta = \pm 1$ has opposite signs on the two checkerboard sublattices of the square lattice. The Zeeman term minimizes the energy of the square-lattice antiferromagnet when the N\'eel order is orthogonal to the magnetic field and so, we take ${\bm B}_Z \cdot {\bm N} = 0$; the ${\bm B}_Z$ term is not essential to the topological and field-theoretic considerations below, but can be important in understanding the experimental role of the applied field, as we will see below.

\begin{figure*}[tb]
\begin{center}
\includegraphics[width=0.90\textwidth]{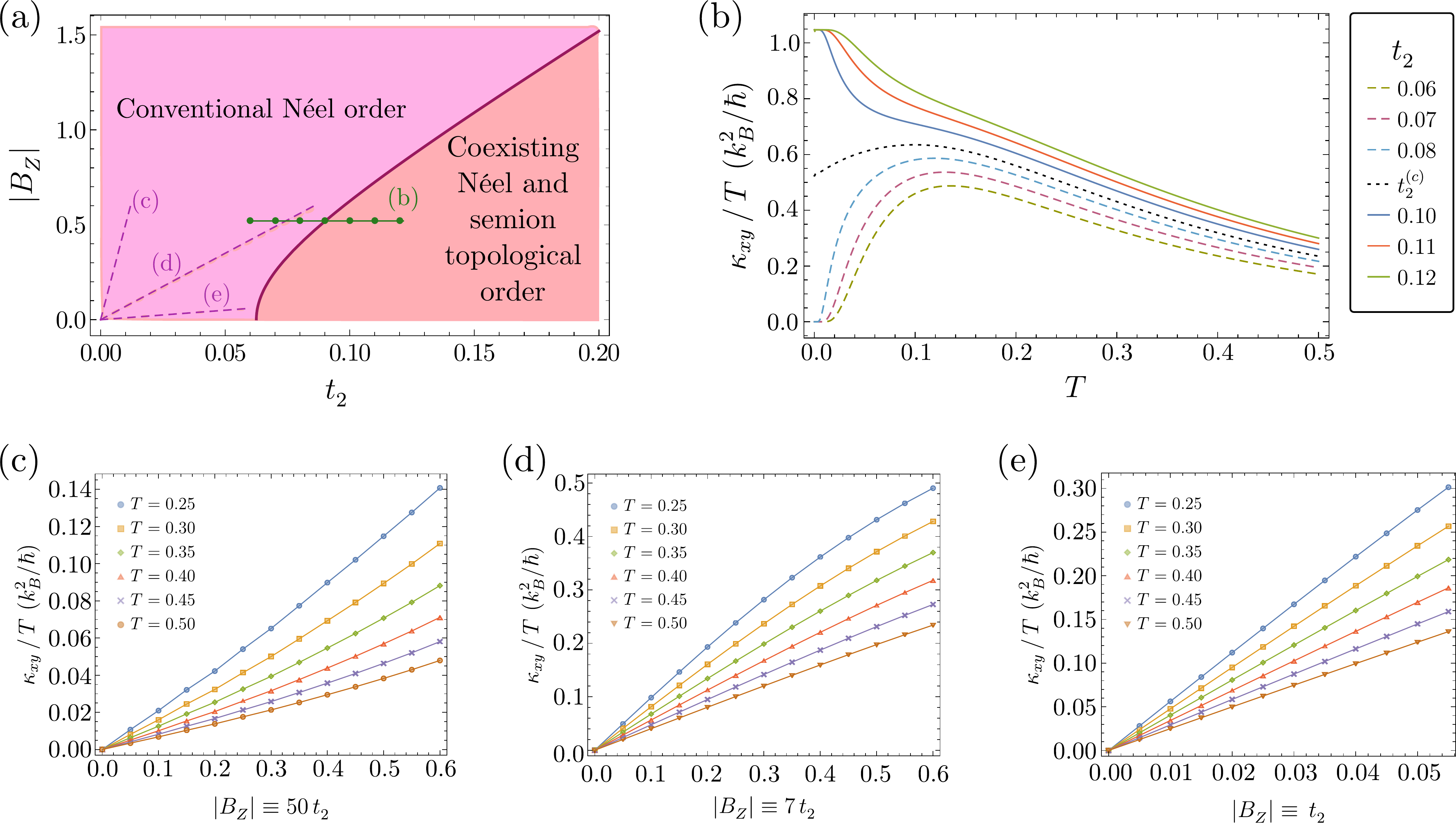}
\caption{\textbf{Phase diagram and thermal Hall conductivity of spinon mean-field theory.} \textbf{a}, The two different phases of the fermionic spinon mean-field Hamiltonian $H_f$ in Eq.~(\ref{eq:Hf}) are shown as a function of the second-nearest-neighbor spinon hopping $it_2$ [see Fig.~\ref{fig:fermionpi}a] and the Zeeman field $|{\bm B}_Z|$. Here, we take the N\'eel order ${\bm N} = 0.5 \hat{z}$ and measure all energies in units of the nearest-neighbor spinon hopping $t_1$. As discussed in the main text, $it_2$ is induced by the orbital coupling of the magnetic field.  
    Both $t_2$ and $|{\bm B}_Z|$ are linear functions of the applied magnetic field, and the dashed purple lines show three possible trajectories for which we plot the field dependence of $\kappa_{xy}$ in (\textbf{c}--\textbf{e}) for different temperatures $T$.
  \textbf{b}, Temperature dependence of the mean-field $\kappa_{xy}$ as $t_2$ is tuned across the phase boundary; the corresponding discrete values of $|\vec{B}_Z|$ and $t_2$ are indicated by green dots in (\textbf{a}). The quantized value of the ordinate in the topological phase is  $\pi/3$, and the bifurcation point as $T \rightarrow 0 $ is at $\pi/6$. Both values are corrected by gauge fluctuations (the exact quantized value in the topological phase is $\pi/6$).}
\label{fig:bifurcation}
\label{fig:kappaH}
 \label{fig:MFPD}
\end{center}
\end{figure*}

Many key results follow from a consideration of the topology of the spinon band structure implied by $H_f$. Our choice of $t^{}_{ij}$ in Fig.~\ref{fig:fermionpi}a and $\eta_i$ leads to a unit cell with two sites. Combined with the spin label $\alpha$, we obtain a total of four spinon bands, which are half-filled. The key discriminant is the net Chern number of the occupied bands. When this is zero, there will be no Chern-Simons term in the theory for gauge fluctuations, leading to confinement and a conventional N\'eel state. On the other hand, when the net Chern number is 2, we obtain a Chern-Simons term and a state with semion topological order (as argued in Ref.~\cite{Wang17}), coexisting with the N\'eel order here, because ${\bm N} \neq 0$; this state has gapped excitations with semionic statistics, along with the conventional spin-wave modes of the N\'eel state. In this manner, we obtain the mean-field phase diagram shown in Fig.~\ref{fig:MFPD}a.

The evolution of the band structure across the phase boundary in Fig.~\ref{fig:MFPD}a is shown in Fig.~\ref{fig:fermionpi}b-e. Note the appearance of two massless Dirac fermions at the critical point. Away from the critical point, these fermions acquire a common Dirac mass, which has opposite signs in the two phases.

Next, we computed the thermal Hall conductivity $\kappa_{xy}$ across the phase boundary in Fig.~\ref{fig:MFPD}a: the results are shown in Fig.~\ref{fig:bifurcation}b. Denoting the Berry curvature for each band $n =1,\ldots, 4$ by $\Omega_{n\mathbf{k}}$, the thermal Hall conductivity is given by  \cite{qin2011energy}
\begin{equation}
    \kappa^{}_{xy} = -\frac{k_B^2}{\hbar T} \int \mathrm{d}\varepsilon\, \varepsilon^2 \sigma_{xy} (\varepsilon) f'(\varepsilon)
    \label{eq:k_xy}
\end{equation}
where $\sigma_{xy} (\varepsilon) = -\int_{\varepsilon_{n \mathbf{k}}< \varepsilon} \mathrm{d}^2 k \,\, \Omega_{n \mathbf{k}}\, /(4 \pi^2)$ is $\hbar/e^2$ times the Hall conductivity, and $f(\varepsilon)$ is the Fermi function. The corresponding Chern number is
\begin{equation}
C_n = \frac{1}{2\pi} \int \mathrm{d}^2 k \,\, \Omega_{n {\bf k}} \in \mathbb{Z}.
\end{equation}
As $T \rightarrow 0 $,
\begin{equation}
    \frac{\kappa^{}_{xy}}{T} = - \frac{\pi k_B^2}{6 \hbar} \sum_{n \, \in \, {\rm filled~bands}} \hspace*{-0.3cm}C_n.
    \label{eq:k_xy0T}
\end{equation}
Consequently, $\kappa_{xy}/T \rightarrow (\pi/3) (k_B^2/\hbar)$ as $T \rightarrow 0$ in the phase with topological order; quantum gauge fluctuations, to be discussed below, will change the prefactor $(\pi/3)$ to the exact quantized value $(\pi/6)$ in this phase. In the other phase, $\kappa_{xy}/T$ varies nonmonotonically as $T$ is lowered, and eventually vanishes as $T \rightarrow 0$ because the occupied bands have opposite Chern numbers. Note the bifurcation in the $T$ dependence at the phase boundary. Exactly on the phase boundary, the present mean-field theory yields $\kappa_{xy}/T \rightarrow (\pi/6) (k_B^2/\hbar)$ as $T \rightarrow 0$: this value is expected to have universal corrections from gauge fluctuations by some nonrational renormalization factor \cite{SS98}.

We plot the field dependence of $\kappa_{xy}$ withing the N\'eel phase in Figs.~\ref{fig:MFPD}c-e, along the dashed lines in Fig.~\ref{fig:kappaH}a. Note that $\kappa_{xy}$ is a nearly linear function of the field, with a slope which is enhanced as we approach the phase boundary to the state with semion topological order.

\section*{Gauge theories and dualities}

\begin{figure}[tb]
    \centering
    \includegraphics[width=\linewidth]{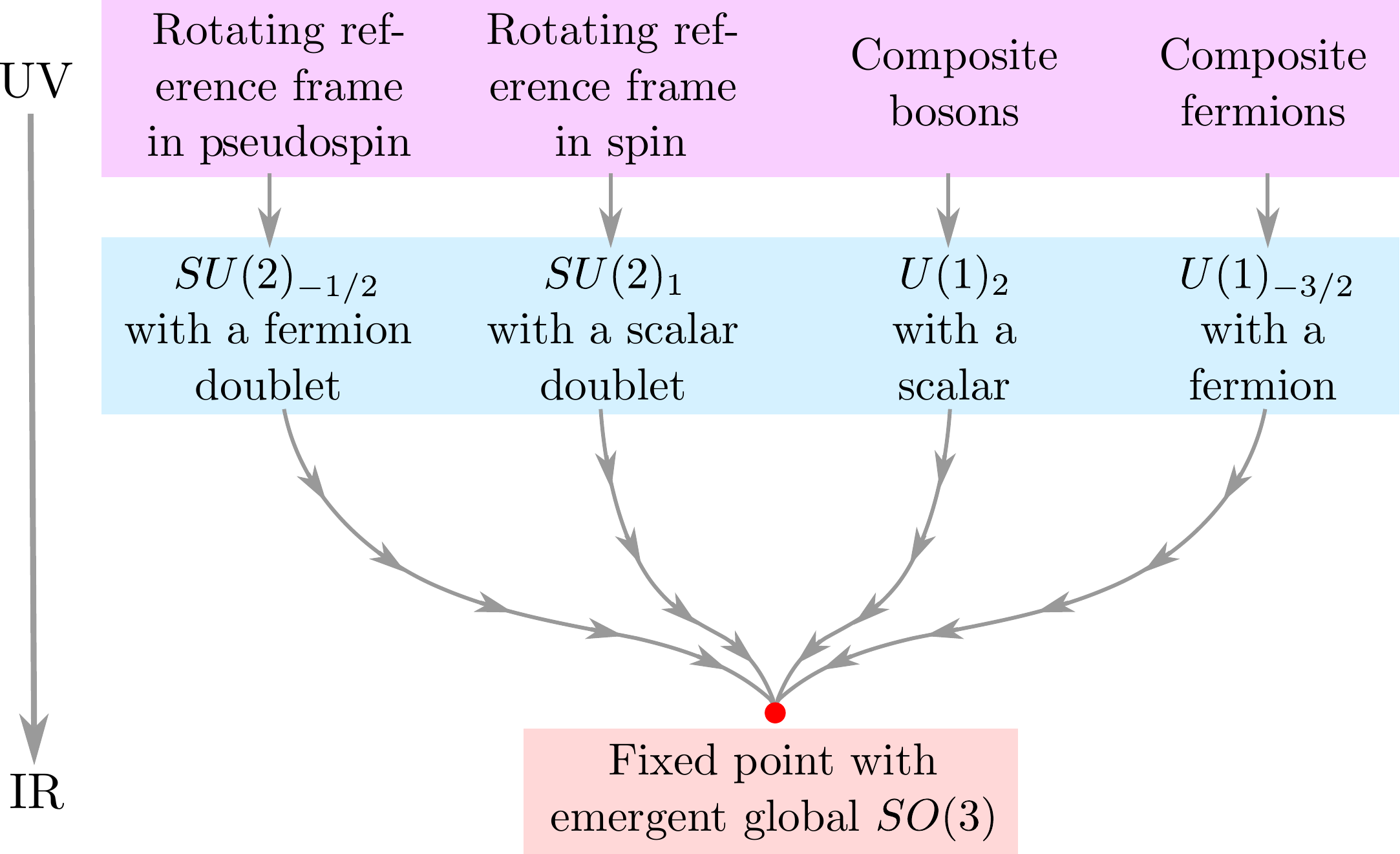}
    \caption{\textbf{Four dual field theories for the antiferromagnet flow to the same fixed point}. Distinct approaches to the lattice antiferromagnet (violet) lead to 4 different continuum field theories (blue) for the transition from the N\'eel to the N\'eel+CSL state (Fig.~\ref{fig:neelvbscsl}). Universality then implies that these describe the same renormalization group fixed point (red) with an emergent global $SO(3)$ symmetry. Adapted from Ref.~\cite{Benini17}.}
    \label{fig:Dualities}
\end{figure}

We now discuss universal properties of the quantum phase transition in Fig.~\ref{fig:MFPD}(a), and the lower phase boundary in Fig.~\ref{fig:neelvbscsl}. 
This critical theory has four different dual formulations, summarized in Fig.~\ref{fig:Dualities}.
The first of these, labeled $SU(2)_{-1/2}$ in Fig.~\ref{fig:Dualities}, is obtained by reinstating gauge fluctuations to the free fermion mean-field theory described above. The resulting field theory turns out to have an emergent global $SO(3)$ symmetry, which must then also be a property of the other dual formulations. We now discuss these field theories, and their connections to the lattice antiferromagnet, in turn. We refer the reader to recent reviews \cite{Witten:2015aba,Seiberg:2016gmd,Seiberg:2016rsg,Cordova:2019jnf} for subtle aspects of gauge and gravitational anomalies which we will not enter into here.

\subsection*{$\bullet$  $SU(2)_{-{1/ 2}}$ with a fermion doublet}
Near the phase boundary in Fig.~\ref{fig:MFPD}a, we can focus on the effective theory of the nearly massless Dirac fermions. These form a single doublet, $\psi$, under the $SU(2)$ gauge symmetry and so, a low energy theory will have a $SU(2)$ gauge field, $A_\mu$, coupled minimally to $\psi$. However, we cannot entirely neglect the single filled fermionic band far from the Fermi level, see Fig.~\ref{fig:fermionpi}b-e. This band has a nonzero Chern number, and integrating out these fermions yields a Chern-Simons term for $A_\mu$ at level $-1/2$. In this manner, we obtain the low-energy 2+1 dimensional Lagrangian
\beq
\mathcal{L}_1 = i \overline{\psi} \gamma^\mu (\partial_\mu - i A_\mu) \psi + m \, \overline{\psi} \psi  - \frac{1}{2} CS [A_\mu]\,. \label{eq:L1}
\eeq
Here, $\gamma^\mu$ are the Dirac matrices, $m$ is the mass term which changes sign across the phase transition, and the last term represents the $SU(2)$ Chern-Simons term. When $m$ is nonzero we can safely integrate the fermions out. For one sign of $m$, the net Chern-Simons term vanishes, and we obtain a `trivial' confining phase with $\kappa_{xy}=0$.
For the other sign of $m$, we obtain a Chern-Simons term at level 1, and the $SU(2)_1$ theory describes a chiral spin liquid with $\kappa_{xy}/T = (\pi/6) (k_B^2 / \hbar)$ \cite{Cappelli:2001mp}.

The emergent global $SO(3)$ symmetry of $\mathcal{L}_1$ is most easily seen by writing $\psi$ in terms of Majorana fermions, and then the fermion kinetic term has a global $O(4)$ symmetry. One $SU(2)$ subgroup of $O(4)$ is the $SU(2)$ gauge symmetry, while the other leads to the global $SO(3)$ symmetry. There is no global $SO(3)$ symmetry in the lattice Hamiltonian $H_f$, so this symmetry is special to the vicinity of the critical point.
Further discussion, including an interpretation of this global symmetry in terms of the microscopic spins, can be found in the Supplementary Information. 

\subsection*{$\bullet$ $U(1)_{2}$ with a charged scalar}
The second dual theory has a complex scalar, $\phi$, coupled to a $U(1)_2$ gauge field, $a_\mu$:
\beq
\mathcal{L}_2 = |(\partial_\mu - i a_\mu) \phi|^2 - s |\phi|^2 - u (|\phi|^2)^2 + \frac{ \epsilon^{\mu\nu\lambda}}{2 \pi} a_\mu \partial_\nu a_\lambda\,. \nonumber
\eeq
However, the $SO(3)$ global symmetry is not manifest in this formulation, and its description requires consideration of monopole operators \cite{Benini17,Aharony15,Aharony16,Hsin16}. The coupling $s$ tunes across the phase transition at $s=s_c$, while the quartic nonlinearity $u$ is assumed to flow to a fixed-point value, analogous to that in the Wilson-Fisher theory without the Chern-Simons term. For $s<s_c$, the $\phi$ field forms a Higgs condensate, and this quenches $a_\mu$ and all topological effects: we thus obtain the conventional N\'eel state.  This maps to the positive-mass phase of the $SU(2)_{-{1\over 2}}$ fermion theory discussed above.
For $s>s_c$, we obtain the state with semion topological order: the gapped $\phi$ quasiparticles have mutual semion statistics which is induced by the Chern-Simons term.  Below the quasiparticle gap, this phase is described by $U(1)_2$ and so, maps to the negative-mass phase of the $SU(2)_{-{1\over 2}}$ fermion theory.
We can connect the field theory $\mathcal{L}_2$ to the lattice antiferromagnet by viewing the latter as a theory of hard-core bosons $S_+ = S_x + i S_y$; then, assuming the bosons form a $\nu=1/2$ fractional quantum Hall state, as in the chiral spin liquid \cite{Kalmeyer}, we identify $\phi$ as the quasiparticle (vortex) operator in the Chern-Simons-Landau-Ginzburg theory \cite{Zhang92,Vishwanath18}.

\subsection*{$\bullet$ $U(1)_{-{3/ 2}}$ with a charged fermion}
The third dual theory of Ref.~\cite{Benini17} is a theory that had been discussed in Ref.~\cite{Barkeshli12}: it has a single Dirac fermion coupled to a $U(1)_{-{3/ 2}}$ gauge field. The $SO(3)$ symmetry is not manifest. Such a field theory can be related to a fractionalization of the hard-core boson $S_+$ into two fermions $\sim f_1 f_2$ \cite{Barkeshli12}. In the state with topological order, which is a $\nu=1/2$ fractional quantum Hall state of the bosons (as above), both fermions fill bands with unit Chern number, as in composite fermion theory \cite{JainPRL}; the phase transition maps to a change of the Chern number of one band to
zero \cite{CFW93,Barkeshli12,Vishwanath18}.

\subsection*{$\bullet$ $SU(2)_{1}$ with a scalar doublet}
The fourth dual theory \cite{Benini17,Aharony15,Aharony16,Hsin16} has a complex scalar doublet transforming as the fundamental of a $SU(2)_1$ gauge field. This can be connected to the $SU(2)$ gauge theory obtained by transforming to a rotating reference frame in spin space \cite{scheurer2018orbital,SS09,SS18}, in which we write the electrons as $c^{}_{\alpha} = R^{}_{\alpha\beta} \psi^{}_\beta$. Here $\psi^{}_\beta$ is a fermion (the `chargon'), and $R^{}_{\alpha\beta}$ is a $SU(2)$ matrix ($R^\dagger R = 1$), which can be expressed in terms of the aforementioned complex scalar doublet; in the renormalized continuum theory, the unit-length constraint on the scalar doublet can be replaced by a quartic self-interaction.
The $SU(2)$ gauge symmetry corresponds to right multiplication of $R$ (and left multiplication of $\psi$),
while the $SO(3)$ global symmetry corresponds to left multiplication of $R$.
We assume that the band structure of the $\psi^{}_\beta$ fermions is such that both species are in a filled band with unit Chern number; then, integrating out these gapped fermions yields the Chern-Simons terms for the $SU(2)_1$ gauge field. Now, the needed transition is obtained by the Higgs transition of the scalar: the topological phase has $R$ gapped, while the trivial phase has $R$ condensed. We also need spectator $SU(2)$ gauge-neutral electrons in filled Chern bands to match the
thermal and electrical Hall conductivities of the two phases.


\section*{Discussion}
For thermal Hall measurements in the cuprates, our main mean-field results are in Fig.~\ref{fig:bifurcation}. Note the large rise in the thermal Hall response in the conventional N\'eel state proximate to the phase boundary, before it eventually vanishes at low enough $T$: this rise is our proposed explanation for the observations of Grissonnanche \textit{et al}.~\cite{Taillefer19}. The field and temperature dependences of $\kappa_{xy}$ in Fig.~\ref{fig:kappaH} match well with observations. It is possible that stronger fields will drive the cuprates across the quantum phase transition into a state with semion topological order, but the stronger field also enhances the Zeeman term, and Fig.~\ref{fig:MFPD}a shows that this term is detrimental to such a transition.

We also discussed gauge-field fluctuation corrections to the results in Fig.~\ref{fig:bifurcation}. We noted that in the topological phase such corrections renormalize the thermal Hall conductivity from $(\pi/3) k_B^2 T/\hbar$ to $(\pi/6) k_B^2 T/\hbar$ as $T \rightarrow 0$. 
Computation of the analogous corrections at higher $T$ and across the phase boundary in Fig.~\ref{fig:MFPD} is more challenging.
The critical theory of the phase boundary was shown to be a central actor in recent studies of dualities of strongly interacting conformal field theories in 2+1 dimensions \cite{Benini17,Aharony15,Aharony16,Hsin16}. The theory of interest has four different formulations which we summarize in \figref{fig:Dualities}; we also provided lattice interpretations of all four field theories in terms of the degrees of freedom of the square-lattice antiferromagnet.
An expansion in the inverse number of matter flavors (analogous to Ref.~\cite{SS98}) is a promising route to computing the universal nonzero temperature thermal Hall effect in these gauge theories near the quantum critical point in Fig.~\ref{fig:bifurcation}a.

Finally, let us comment on the role of fluctuations of the N\'eel order parameter. Spin waves make only a small contribution to the thermal Hall effect \cite{SCSS19}. In two spatial dimensions, thermal fluctuations of the N\'eel order restore spin rotation symmetry at all nonzero $T$ \cite{CHN89}, but these classical fluctuations are not expected to significantly modify the quantum criticality of the topological quantum phase transition in Fig.~\ref{fig:MFPD}a, which involves no change in symmetry.

\section{Methods}


\textbf{Deconfined criticality and phase diagram.} The square-lattice antiferromagnet with first ($J_1$) and second ($J_2$) neighbor exchange interactions has been the focus of many numerical studies in the past decades. There appears to be general agreement that increasing $J_2/J_1$ destroys the N\'eel state and leads to a state with valence bond solid (VBS) order \cite{Sheng18a,Sandvik18}. There is also significant evidence that the transition region between these states is described by a deconfined critical field theory \cite{Senthil04} over a large intermediate length scale \cite{Sheng18a}. Furthermore, there is strong support for a global $SO(5)$ symmetry between the N\'eel and VBS orders \cite{Nahum18} over this scaling region, as is expected for the deconfined critical theory \cite{Tanaka05,Senthil06,Wang17}.
The ultimate fate of the phase transition at the longest distances remains unsettled, but it is plausible that it is described by a complex fixed point, very close to the real physical axis \cite{Wang17,Rychkov18,Ma:2018euv}.

It is useful to now consider the phase diagram of the $J_1$-$J_2$-$J_\chi$ antiferromagnet on the square lattice by starting from a theory in which the $SO(5)$ symmetry is initially explicit. This is just the fermionic spinon representation used in Eq.~(\ref{eq:fullH}). In the absence of N\'eel order (${\bm N} = 0$) and an applied field (${\bm B}_Z = 0$), the continuum limit of Eq.~(\ref{eq:fullH}) yields two flavors of two-component Dirac fermions, which are then coupled to a $SU(2)$ gauge field; so instead of Eq.~(\ref{eq:L1}) we now have \cite{Wang17}
\beq
\mathcal{L}_{SO(5)} = i \overline{\psi}_a \gamma^\mu (\partial_\mu - i A_\mu) \psi^\pdagger_a + m_\chi \overline{\psi}_a \psi^\pdagger_a
\eeq
where $a=1,2$ is the flavor index. 

The $SO(5)$ symmetry is apparent after we express $\mathcal{L}_{SO(5)}$ in terms of Majorana fermions. The fermion mass $m_\chi \propto J_\chi$ is also $SO(5)$ invariant, and is a perturbation on the putative $SO(5)$-invariant N\'eel-VBS critical point at $m_\chi = 0$. It is plausible that $m_\chi$ is a relevant perturbation on such a critical point (with scaling dimension $\lambda_\chi > 0$), and then an {\it infinitesimal\/} $J_\chi$ will be sufficient to drive the critical antiferromagnet into the chiral spin liquid phase. Should the N\'eel--VBS transition be weakly first-order, then a very small value of $J_\chi$ will be sufficient. Tuning away from the critical point by changing the value of $J_2/J_1$ yields a second relevant perturbation to the critical point (with scaling dimension $\lambda_2 > 0$) which explicitly breaks $SO(5)$ symmetry, but is allowed by the symmetries of the underlying antiferromagnet. We obtain the phase diagram proposed in Fig.~\ref{fig:neelvbscsl}
upon considering the interplay of these perturbations; all phase boundaries scale as $J_\chi \sim |J_2 - J_{2c}|^{\lambda_\chi/\lambda_2}$ for a $SO(5)$ critical point at $J_\chi=0$, $J_2 = J_{2c}$, so for $\lambda_\chi > \lambda_2$, we obtain the onset of semion topological order at small values of $J_\chi$ even away from the $SO(5)$ point. In the limit of a large number of fermion flavors, $\lambda_\chi =1$ and $\lambda_2 = -1$, and $\lambda_\chi > \lambda_2$ is thus plausible.
This phase diagram is compatible with the small system size studies of Ref.~\cite{Cirac13}.
We emphasize that the existence of a $SO(5)$ critical point is not a requirement for the existence of a continuous N\'eel to N\'eel+CSL transition with SO(3) symmetry described in the main part of the paper.

\section{Data availability}
The data that support the plots within this paper and other findings of this study are available from the corresponding author upon reasonable request.

\bibliographystyle{apsrev4-1_custom}
\bibliography{THF_2}

\section{Acknowledgements}
\begin{acknowledgments}
This research was supported by the National Science Foundation under Grant No.~DMR-1664842. SC acknowledges support from the ERC synergy grant UQUAM. MS acknowledges support from the German National Academy of Sciences Leopoldina through grant LPDS 2016-12. We thank N. Seiberg for explaining many subtle aspects of the non-Abelian dualities to us. We thank G.~Grissonnanche, Yin-Chen He, C.~Hickey, Chao-Ming Jian, P. A. Lee, A.~Nahum, L.~Taillefer, and Liujun Zou for helpful discussions.
\end{acknowledgments}

\section{Author contributions}
All authors contributed to the research leading to the formulation and analyses of the quantum field theory, and the writing of the paper.
R.S. performed the numerical mean-field computations presented in Figs 2 and 3.

\section{Competing interests}
The authors declare no competing interests.

\newpage

\onecolumngrid
\setcounter{equation}{0}\setcounter{page}{1}
\linespread{1.25}
\setlength{\parskip}{1mm plus 0.5mm minus 0mm}

\section{Supplementary Information for \\ Enhanced thermal Hall effect in the square-lattice N\'eel state}

\begin{center}
Rhine Samajdar, Mathias S.~Scheurer, Shubhayu Chatterjee, Haoyu Guo, Cenke Xu, and Subir Sachdev

\end{center}

\subsection{1. $SO(3)$ symmetry on the lattice}
We investigate the emergent global $SO(3)$ symmetry at the critical point by explicitly constructing an order parameter that transforms as a vector under this $SO(3)$ symmetry in terms of the low-energy fermionic degrees of freedom, starting with the lattice-scale Hamiltonian $H_f$ in Eq.~(3). For simplicity and to obtain a transparent physical picture of this vector order parameter, we first set ${\bm B}_Z = 0$ in $H_f$, and later discuss the generalization to ${\bm B}_Z \neq 0$.

The hopping part of $H_f$ (obtained by setting ${\bm N} = 0 = {\bm B}_Z$) describes a chiral spin liquid, and is symmetric under square-lattice translations $T_{i}$ (with $i = x,y)$, $C_4$ rotations about a site, time-reversal accompanied by reflections  $\Theta R_i$, and global SU(2) spin-rotations \cite{scheurer2018orbital}. Turning on a nonzero N\'eel order parameter ${\bm N} = N \hat{z}$ $(N \neq 0)$ breaks time-reversal, spin-rotation and translation symmetries explicitly, and results in a reduced set of symmetries generated by: 
\beq
T_{x+y}&:& \S_{(i_x,i_y)} \rightarrow \S_{(i_x+1,i_y+1)} \nn
C_4&:& \S_{(i_x,i_y)} \rightarrow \S_{(-i_y,i_x)} \nn
\Theta R_x &:& \S_{(i_x,i_y)} \rightarrow (-S^1,S^2,S^3)_{(-i_x,i_y)}  \nn
U_s(\hat{z},\theta) &:& \S_{(i_x,i_y)} \rightarrow e^{i \sigma^3 \theta/2} \S_{(i_x,i_y)} \, e^{-i \sigma^3 \theta/2} \nn
\Sigma_x \equiv U_s(\hat{x}, \pi) T_x &:& \S_{(i_x,i_y)} \rightarrow e^{i \sigma^1 \pi/2} \S_{(i_x+1,i_y)} e^{-i \sigma^1 \pi/2}
\label{eq:Sym}
\eeq
In Eq.~(\ref{eq:Sym}), $U_s(\hat{z},\theta)$ is the remaining U(1) spin-rotation symmetry about the z-axis, and $\Sigma_x $ denotes a spin-rotation about an axis perpendicular to the N\'eel vector [chosen here to be $U_s(\hat{x},\pi) = i \sigma^1$ w.l.o.g.], accompanied by unit lattice translation $T_x$. Note that $T_{-x+y} = T_{x+y}(\Sigma_x)^{-2}$, $\Theta R_y = U_s(\hat{z},\pi) C_4^2 (\Theta R_x)^{-1}$ and $\Sigma_y = \Sigma_x T_{-x+y}$ are not independent symmetry generators. Further, note that $H_f$ is invariant under any lattice symmetry operation $S$ only up to an additional gauge transformation $f_{i\sigma} \rightarrow e^{i \phi_S [S(i)]} f_{S(i),\sigma}$; in other words, the lattice symmetries act projectively and the combination of $S$ and $e^{i \phi_S} $ leaves $H_f$ invariant. For the set of symmetries in Eq.~(\ref{eq:Sym}), the required gauge-transformations are $\phi_{T_{x+y}} = 0 = \phi_{\Theta R_x}$, while $\phi_{C_4} =  \pi(i_x-i_y)(i_x-i_y+1)/2$ and $\phi_{\Sigma_x} = \pi \, i_y$.

Now, we turn to diagonalizing $H_f$ in Eq.~(3) with ${\bm B}_Z = 0$ and finding the relevant low-energy fermionic modes. Defining a two-site unit cell along the $\hat{x}$-direction with sublattices A (even parity of $i_x+i_y$) and B (odd parity of $i_x+i_y$), the Hamiltonian from Eq.~(3) is given in momentum space by (setting the lattice spacing $a=1$): 
\begin{alignat}{1}
\nonumber H &= \sum_{\k} f^\dagger_\k h_\k^\pdagger f_\k^\pdagger; \quad f_\k = \left( f^\pdagger_{\k A \ua} \,\, f^\pdagger_{\k B \ua} \,\, f^\pdagger_{\k A \da} \,\, f^\pdagger_{\k B \da}\right)^T, \\
h^\pdagger_\k &= \begin{pmatrix} \tilde{h}_{\k}(N) & 0 \\ 0 & \tilde{h}_{\k}(-N) \end{pmatrix}, \;   \\
\nonumber \tilde{h}^\pdagger_{\k}(N)  &= \begin{pmatrix} -\frac{N}{2} - 4 t_2 \sin k_x \cos k_y & - 2 t_1 (\cos k_x + i \sin k_y) \\
- 2 t_1 (\cos k_x - i \sin  k_y) & \frac{N}{2} + 4 t_2 \sin k_x \cos k_y
\end{pmatrix}.
\end{alignat}
The critical point, with Dirac cones at $\pm \Q$, appears at $N = \pm 8 t_2$. In our gauge choice, it holds $\Q = (\pi/2,0)$. For concreteness let us focus on the critical point at $N = -8 t_2$. Then, the low-energy bands at valley $\Q$ are entirely in the spin-up sector; the remaining (spin-down) bands can be integrated out without renormalizing the low-energy bands. A similar picture holds for the valley $- \Q$, where the low-energy bands are tied to the spin-down sector, as shown by the linearized Hamiltonian near $-\Q$. Therefore, the effective low-energy Hamiltonian is given by:
\beq
H^\pdagger_{\textrm{eff}} = 2 t^\pdagger_1 \sum_{q \leq \Lambda} \psi^\dagger_{\q, a} \left( q^\pdagger_x \tau^1 + q^\pdagger_y \tau^2 \right) \psi^\pdagger_{\q, a}
\eeq
where $\Lambda$ is some UV momentum cutoff, the matrices $\tau^\alpha$ act in sublattice space, and
\beq
\psi_{\q, a} = \begin{cases} \begin{pmatrix} f_{\Q + \q, A, \ua} \\ f_{\Q + \q, B, \ua} \end{pmatrix} \text{ for } a = (\Q, \ua), \\ -i\tau^2 \begin{pmatrix} f_{-\Q +\q, A, \da} \\ f_{-\Q +\q, B, \da} \end{pmatrix} \text{ for } a = (-\Q, \da). \end{cases}
\eeq
For our analysis, it is useful to revert to real space and define Nambu spinors $C_{A}(\r)$ and $C_{B}(\r)$ that vary slowly on the lattice scale by projecting the lattice fermions $f_{i\sigma}$ onto the low-energy bands: 
\begin{eqnarray}
\label{eq:RealSpinors}
 \begin{pmatrix} f^\dagger_{i A \da} \\ f^\pdagger_{i A \ua} \end{pmatrix} \approx \begin{pmatrix} f^\dagger_{- A }(\r) \\ f^\pdagger_{+ A }(\r) \end{pmatrix} e^{i \Q \cdot \r_i} \equiv C^\pdagger_{A}(\r) e^{i \Q \cdot \r_i}, \\
\nonumber 
 \begin{pmatrix} f^\dagger_{i B \da} \\ f^\pdagger_{i B \ua} \end{pmatrix} \approx \begin{pmatrix} i f^\dagger_{- B }(\r) \\ i f^\pdagger_{+ B }(\r) \end{pmatrix} e^{i \Q \cdot \r_i} \equiv  i \, C^\pdagger_{B}(\r) e^{i \Q \cdot \r_i}.
\end{eqnarray}
In Eq.~(\ref{eq:RealSpinors}) and henceforth, we will denote each two-site unit cell by $i$, which is also the location of the basis site A (hence $\eta_i = 1$). We perform a further redefinition of the low-energy field operator in order to make contact with subsection 2 on the continuum approach (which follows the conventions of Ref.~\cite{Wang17}). 
\beq
\Psi_{\alpha ,a}(\r) = \left( e^{i \frac{\pi}{4} \tau^1} e^{i \frac{\pi}{4}\tau^3}\right)_{\alpha s} (V^s)_{a b} ~ C_{s,b}~, \text{ with } V^A  = i \eta^3 \text{ and } V^B = \eta^0
\eeq
where $s = A$ or $B$ denotes sublattice, $a$ or $b$ $(=\pm)$  indexes the Nambu/valley space, and all repeated indices are summed over. Upon re-expressing the degrees of freedom in terms of $\Psi_{\alpha, a}(\r)$, the low-energy Lagrangian takes the following form [to which gauge fluctuations $A_\mu$ can be readily introduced, see Eq.~(7) of the main text]:
\beq
\mathcal{L}_{\textrm{eff}} =  i \int d^2\r \; \bar{\Psi}_{\alpha,a}(\r) \left( \gamma^\mu \right)_{\alpha \beta} \partial_\mu \Psi_{\beta,a}(\r),
\label{eq:Leff}
\eeq
where we have re-scaled spacetime to set the Dirac velocity $v_F = 2t_1 = 1$, defined $\bar{\Psi} \equiv \Psi^\dagger \gamma^0$ and the Dirac Gamma matrices are given by $(\gamma^0, \gamma^x, \gamma^y) = (\tau^2, i \tau^3, i \tau^1)$. 

 One crucial point to note is that the antiunitary time-reversal symmetry $\Theta$ (which flips both spin and valley) re-emerges in the low-energy action $L_{\textrm{eff}}$ in Eq.~(\ref{eq:Leff}), although it is broken explicitly in the lattice Hamiltonian $H_f$ in Eq.~(3) as
\beq
&& \Theta: f_{\pm, s} \rightarrow \pm f_{\mp,s}, \text{ with } \Theta^2 = -1. 
\eeq
Intuitively, this happens because time-reversal simultaneously flips spin and valley; $\mathcal{L}_{\textrm{eff}}$ is invariant under such a transformation. The effect of $t_2$ (that describes a local orbital magnetic field) and the N\'eel order parameter ${\bm N}$ (that acts as a local Zeeman field) thus counteract each other in the low-energy effective field theory.  

In order to make the emergent SO(3) symmetry manifest in the low-energy action, it is useful to construct the following matrix of low-energy spinors, in analogy with Ref.~\onlinecite{Wang17}:
\beq
X_\alpha(\r) = \begin{pmatrix} \psi^\pdagger_{\alpha, + \ua}(\r)   & - \psi^\dagger_{\alpha,-\da}(\r)  \\ \psi^\pdagger_{\alpha,-\da}(\r)  & \psi^\dagger_{\alpha,+\ua}(\r)  \end{pmatrix}  \text{ where } \Psi_{\alpha} \equiv \begin{pmatrix} \psi^\dagger_{\alpha,-\da} \\ \psi_{\alpha,+\ua}  \end{pmatrix}
\eeq
The low-energy gauge-fluctuations act on $X_\alpha(\r)$ via right multiplications, as follows:
\beq
SU(2)_g: X_\alpha(\r) \rightarrow X_\alpha(\r) \, U_g^\dagger(\r)
\eeq
The effective action in Eq.~(\ref{eq:Leff}), including gauge fluctuations $A_\mu$, can be recast in terms of $X_\alpha(\r)$ (repeated indices $\alpha$ and $\mu$ are summed over):
\beq
\mathcal{L}_{\textrm{eff}} = \frac{i}{2} \int d^2\r \Tr [ \bar{X}_\alpha(\r) \gamma^\mu D^A_\mu X_\alpha(\r) ], \text{ where } D^A_\mu X_\alpha = \partial_\mu X_\alpha + i X_\alpha A_\mu \text{ and }  \bar{X}_\alpha =  X^\dagger_\alpha \gamma^0
\label{eq:MatAct}
\eeq
The action in Eq.~(\ref{eq:MatAct}) is clearly invariant under an emergent global SU(2) symmetry $U$ that acts on $X_\alpha$ by left multiplication:
\begin{alignat}{1} 
SU(2)_{\text{sym}}: X_\alpha(\r) \rightarrow U X_\alpha(\r)
\end{alignat}
Therefore, the manifestly gauge-invariant order parameter that is constructed entirely out of the low-energy degrees of freedom, and transforms as a vector under the emergent global $SO(3)$ is given by
\beq
O^a(\r) \equiv \sum_{\alpha} \Tr\left[X_\alpha^\dagger(\r) \sigma^a X_\alpha(\r) \right],
\eeq
where $\sigma^a$ are the Pauli matrices in the mixed spin-valley space labeled by $a = \{1,2,3\}$. The symmetry transformations of $O^a$ can be worked out by considering their (projective) action on the real space lattice-scale spinors in Eq.~(\ref{eq:RealSpinors}), and subsequently deducing their action on $\psi_{\alpha,a}(\r)$ and hence on $O^a(\r)$. The results are presented in Table \ref{tab:SymTrans}. 

\begin{center}
\begin{table*}
    \centering
    \begin{tabular}{c cc | c cc cc cc cc cc cc cc cc cc}
    \hline\hline
         &{Observable} && & $T_{x+y}$   & &  $T_{-x+y}$ & & $\phantom{-}C_4$ & & $\phantom{-}\Sigma_x$  & & $\phantom{-}\Sigma_y$ && $\Theta R_x$ && $\Theta R_y$ && $U_s(\hat{z},\theta)$ && $\phantom{-}\Theta$ & \\[5pt]  \hline
         & $O^1$ && & $- O^1$ && $- O^1$ && $-O^2$ && $\phantom{-}O^1$ && $-O^1$ && $-O^1$ && $-O^1$ && $\cos\theta \; O^1 + \sin\theta \; O^2$ && $\phantom{-}O^1$ & \\[4pt] 
         & $O^2$ && & $- O^2$ && $- O^2$ && $\phantom{-}O^1$ && $ -O^2$ && $\phantom{-}O^2$ && $\phantom{-}O^2$ && $\phantom{-}O^2$ && $-\sin\theta \; O^1 + \cos\theta \; O^2$ && $\phantom{-}O^2$ & \\[4pt] 
         & $O^3$ && & $\phantom{-}O^3$ && $\phantom{-}O^3$ && $\phantom{-}O^3$ && $- O^3$ && $-O^3$ && $\phantom{-} O^3$ && $\phantom{-} O^3$ && $\phantom{-}O^3$ && $-O^3$ &
        \\ \hline\hline
    \end{tabular}
    \caption{Transformation of the components of the gauge-invariant fermion-bilinear $SO(3)$ vector under the symmetries of $H_f$ (for ${\bm B}_Z = 0$) and the emergent time-reversal symmetry $\Theta$.}
    \label{tab:SymTrans}
\end{table*}
\end{center}

\begin{figure*}[t]
    \centering
    \includegraphics[scale=0.5]{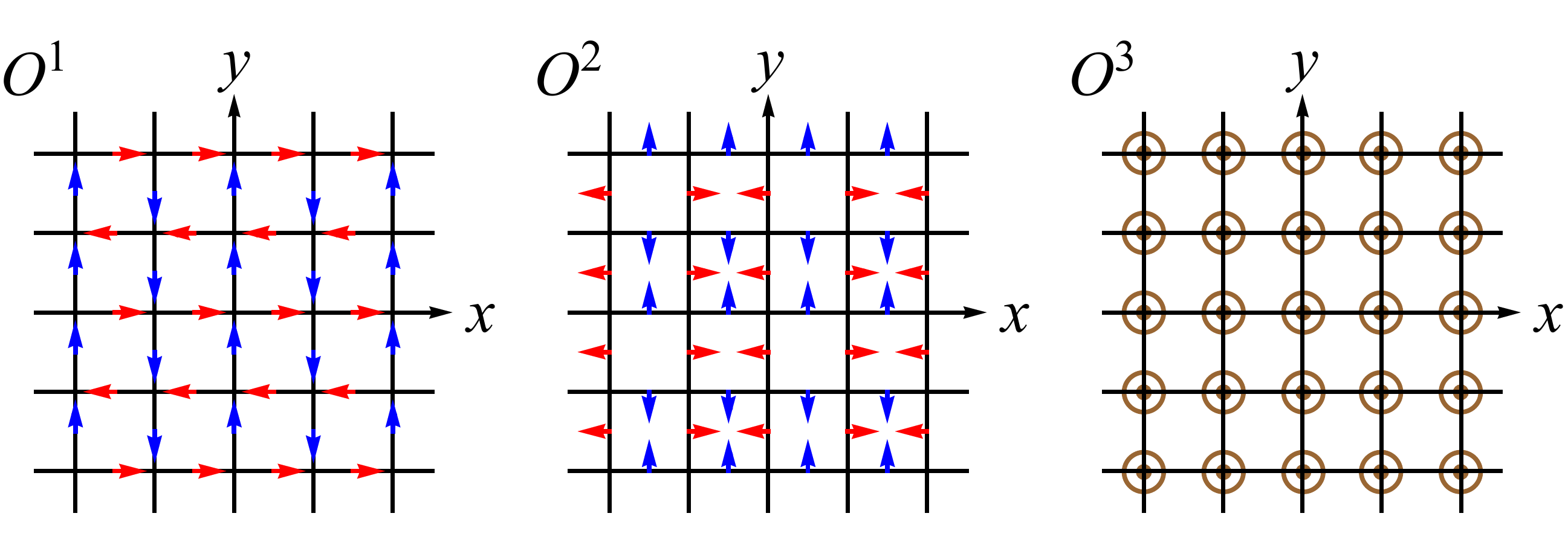}
    \caption{Pictorial representation of the components of the $SO(3)$ vector $O^a$ defined in \equref{OaOperators}. For $O^1$ and $O^2$, we plot the corresponding spin-chirality vector ${\bm S}_i \times {\bm S}_{j}$ on each bond $\langle ij \rangle$. For $O^3$, we plot $(\bm S_i)_3$ on each lattice site $\r_i$.}
    \label{fig:OPfig}
\end{figure*}

Next, we construct a representation of $O^a$ in terms of the microscopic spin operators ${\bm S}_i$. The rotation of $O^2$ into $O^1$ under either a $C_4$ rotation or a  spin-rotation $U_s(\hat{z},-\pi/2)$ indicates that these must be spin-orbit coupled. Indeed, one can check that the following representatives satisfy all the symmetry constraints in Table~\ref{tab:SymTrans} (the indices $1,2$ represent directions in spin-space, while $x,y$ represent directions in real-space, and $N_s$ is the number of lattice sites). 
\beq
O^1 & = & \frac{1}{N_s} \sum_{i} (-1)^{i_x} ({\bm S}_i \times {\bm S}_{i + \hat{y}})_2 + (-1)^{i_y} ({\bm S}_i \times {\bm S}_{i + \hat{x}})_1 \nn
O^2 & = & \frac{1}{N_s} \sum_{i}  (-1)^{i_y} ({\bm S}_i \times {\bm S}_{i + \hat{x}})_2 - (-1)^{i_x} ({\bm S}_i \times {\bm S}_{i + \hat{y}})_1 \nn
O^3 & = & \frac{1}{N_s} \sum_{i} ({\bm S}_i)_3 \label{OaOperators}
\eeq
Thus, $O^1$ and $O^2$ can be interpreted as bond-direction dependent spin-projections of a vector-spin chirality operator ${\bm S}_i \times {\bm S}_{j}$, which is even under time-reversal $\Theta$; related operators were discussed in Ref.~\cite{LuRan10}. The final component $O^3$ is unchanged under the action of $U_s(\hat{z},\theta)$, but is odd under unit lattice translations accompanied by spin-flip and also under time-reversal; therefore, it can be interpreted as an Ising ferromagnetic order parameter. The operators in \equref{OaOperators} are illustrated graphically in \figref{fig:OPfig}.

Lastly, we comment on the addition of an external Zeeman field ${\bm B}_Z$ perpendicular to the N\'eel order parameter ${\bm N}$, which gets rid of the residual U(1) spin-rotation symmetry generated by $S^z$ in $H_f$. In this case, the only leftover symmetry generators at the lattice scale are $T_{x+y}$, $C_4$, $\Theta R_y$ and $\Sigma_x$ (for ${\bm B}_Z = B \, \hat{x}$, $\Sigma_i = U_s(\hat{x},\pi) T_i$ remains a symmetry for each $i = x,y$). In the corresponding low-energy action $\mathcal{L}_{\text{eff}}$, a global $SO(3)$ and time-reversal $\Theta$ emerge as symmetries just as in the previous case. Hence, one can label the low-energy fermions by their valley indices ($v = +\Q$ or $-\Q$) and carry out an analogous construction of a gauge-invariant order parameter $O^a(\r)$ that transforms as a vector under this emergent $SO(3)$ symmetry. $O^i$ ($i = 1,2$) break $T_{x+y}$, $\Sigma_{\bar{i}}$ and $C_4$ (where $\bar{i} = y$ for $i = x$ and vice-versa). While $O^3$ is invariant under all lattice symmetries that do not involve spin-flips, it breaks both $\Sigma_i$ (for $i = x,y$) and time-reversal $\Theta$, which is emergent at the critical point. We conclude that the results in Table \ref{tab:SymTrans} are still valid for ${\bm B}_Z \neq 0$, modulo the columns $U_s(\hat{z},\theta)$ and $\Theta R_x$. The lattice symmetry $C_4$ and $\Sigma_x$ (or $\Sigma_y$) symmetry together can be embedded into group $O(2) = SO(2) \rtimes Z_2$ (the explicit symmetry of two of the four dual field theory descriptions in the main text), which is a subgroup of $SO(3)$. Eventually we expect that the symmetry in the infrared limit is enlarged to $SO(3)$.

\subsection{2. $SO(3)$ symmetry generators in the continuum} 

Another way to investigate the nature of the global $SO(3)$ symmetry is to connect its generators to those of the $SO(5)$ symmetry of the $\pi$-flux state described in detail in Ref.~\cite{Wang17}.
The N\'eel and valence bond solid (VBS) order parameters combine to form a fundamental 5-vector of this $SO(5)$ symmetry. Note that in the considerations here, we are not assuming the existence of a conformal field theory with $SO(5)$ symmetry describing the N\'eel-VBS critical point; the emergent $SO(5)$ symmetry is already a property of the mean-field $\pi$-flux state, and that is sufficient for our purposes.

The $SO(5)$ symmetry is most explicit in the Majorana fermion basis, and the Majorana fermion $\chi$ is
introduced as $\psi = \chi^{}_1 + i \chi^{}_2$. We choose a basis such that the Dirac matrix $\gamma^0 = \sigma^2$; the spin
chirality then becomes $\bar{\chi}\chi = \chi^t \gamma^0 \chi$. Then, in
addition to the Dirac index, the Majorana fermion still carries
three extra indices; each represents a two-component space, which
in total hosts a $SO(8)$ transformation that includes both the
$SU(2)$ gauge and the $SO(5)$ global symmetries. After choosing an
appropriate basis \cite{Wang17}, the generators of the $SU(2)$ gauge group are
\beq G_1 = \sigma^{230}, \ \ G_2 = \sigma^{210}, \ \ G_3 =
\sigma^{020}; \eeq the generators of the $SU(2)$ spin group are \beq
S^1 = \sigma^{120}, \ \ S^2 = \sigma^{200}, \ \ S^3 =
\sigma^{320}. \eeq Here, $\sigma^{abc} = \sigma^a \otimes \sigma^b
\otimes \sigma^c$, and $\sigma^0 = \mathbf{1}_{2\times 2}$. The
N\'eel and valence bond solid (VBS) operators form an $SO(5)$ vector
\begin{alignat}{1}
\mathcal{N}_a &=
\bar{\chi} \Gamma_a \chi\,, \\
\Gamma_1 &= \sigma^{322}, \
\Gamma_2 = \sigma^{122}, \ \Gamma_3 = \sigma^{202}, \ \Gamma_4 =
\sigma^{003}, \ \Gamma_5 = \sigma^{001}\,.
\nonumber
\end{alignat}
One may check that
$[G^a, S^b] = 0$, and $[G^a, \Gamma^b] = 0$ for all $a, b$. The
ten $SO(5)$ generators can be obtained by the commutators of the
five Gamma matrices: $\Gamma_{ab} = \frac{1}{2i}[\Gamma_a,
\Gamma_b]$. If we turn on a background N\'eel order parameter, say
$\langle \bar{\chi}\Gamma^3 \chi \rangle \neq 0$, the $SO(5)$
symmetry is broken down to $SO(4)$, which is generated by
six out of the ten generators of $SO(5)$: \beq \Gamma_{14} =
\sigma^{321}, \ \ \Gamma_{15} = - \sigma^{323}, \ \ \Gamma_{45} =
\sigma^{002}, \cr\cr \Gamma_{24} = \sigma^{121}, \ \ \Gamma_{25} =
- \sigma^{123}, \ \ \Gamma_{12} = \sigma^{200}. \eeq We can
construct two sets of independent $SU(2)$ generators out of the six
generators:
\begin{eqnarray}
T_{A1} &=& \frac{1}{2} (\Gamma_{15} +
\Gamma_{24}), \ T_{A2} = \frac{1}{2} (- \Gamma_{14} +
\Gamma_{25}), \nonumber \\
T_{A3} &=& \frac{1}{2} (\Gamma_{12} +
\Gamma_{45}); \nonumber \\
T_{B1} &=& \frac{1}{2} (\Gamma_{15} -
\Gamma_{24}), \ T_{B2} = \frac{1}{2} (\Gamma_{14} + \Gamma_{25}),
\nonumber \\
 T_{B3} &=& \frac{1}{2} (\Gamma_{12} - \Gamma_{45}).
\end{eqnarray}
Now,
let us turn on two fermion mass terms simultaneously:
$m_1 \bar{\chi}\Gamma_3\chi + m_2 \bar{\chi}\chi$. Then, the $T_{Ai}$ ($T_{Bi}$) are
the needed $SO(3)$ generators that operate on the massless fermion
subspace when $m_1 = - m_2$ ($m_1 = m_2$). We have not included the influence of the Zeeman term here, which has a weak effect near the $SO(5)$-symmetric point.


\end{document}